\documentclass[conference]{IEEEtran}
\IEEEoverridecommandlockouts
\usepackage[nottoc]{tocbibind}
\usepackage{cite}
\usepackage{amsmath}

\usepackage{amsmath,amssymb,amsfonts}
\usepackage{algorithmic}
\usepackage{graphicx}
\usepackage{textcomp}
\usepackage{xcolor}
\usepackage{array}
\usepackage{caption}
\usepackage{subcaption}
\def\BibTeX{{\rm B\kern-.05em{\sc i\kern-.025em b}\kern-.08em
    T\kern-.1667em\lower.7ex\hbox{E}\kern-.125emX}}
\begin{document}

\title{A Matched-filter based method in the Synthetic Aperture Radar Images Using FMCW radar
}

\author{\IEEEauthorblockN{1\textsuperscript{st} Moein Movafagh}
\IEEEauthorblockA{\textit{Dept.of electrical engineering} \\
\textit{K. N. Toosi University of Technology}\\
Tehran, Iran \\
m.movafagh@kntu.ac.ir}
\and
\IEEEauthorblockN{2\textsuperscript{nd} Avik Santra}
\IEEEauthorblockA{\textit{} \\
\textit{Infineon Technologies AG}\\
Neubiberg, Germany \\
avik.santra@infineon.com}
\and
\IEEEauthorblockN{3\textsuperscript{rd} Daniel Oloumi}
\IEEEauthorblockA{\textit{} \\
\textit{Infineon Technologies AG}\\
Villach, Austria \\
Daniel.oloumi@infineon.com}
}

\maketitle

\begin{abstract}
The stretch processing architecture is commonly used for frequency modulated continuous wave (FMCW) radar due to its inexpensive hardware, low sampling rate, and simple architecture. However, the stretch processing architecture is not able to achieve optimal Signal to Noise ratio (SNR) in comparison to the matched-filter architecture. In this paper, we aim to propose a method whereby stretch processing can achieve optimal SNR. Hence, we develop a novel processing method to enable applying a matched filter to the output of the stretch processing. The proposed architecture achieves optimal SNR while it can operate on a low sampling rate. In addition, the combination of the proposed radar architecture and SAR technique can generate high-quality images. To evaluate the performance of the proposed architecture, four scenarios are considered. Simulation is carried out based on these scenarios. The simulation results show that the proposed radar demonstrates the ability to generate an image with higher quality over stretch processing. This proposed radar can also bring a bigger gain compression.        
\end{abstract}

\begin{IEEEkeywords}
FMCW radar, SAR, Matched-filter architecture ,stretch processing architecture
\end{IEEEkeywords}

\section{Introduction}
\IEEEPARstart{R}{ECENTLY}, the frequency modulated continuous wave (FMCW) radars draw attention due to their advantages such as low sampling rate and the minimum target range \cite{ting2017fmcw}. The FMCW radar can be a reasonable alternative for the pulsed radar since it requires a lower transmitting power\cite{meta}. Besides, simplicity, low cost, and miniaturized system design are potential advantages that stimulate demands for FMCW radar in many applications \cite{charvat2006low}. Therefore, the FMCW radar is an ideal choice for both industrial and academic purposes.\par
The combination of FMCW radar and synthetic aperture radar (SAR) principle presents a high-resolution imaging\cite{ting2017fmcw}. The FMCW radar provides a high-range resolution. So, it enables the creation of high-resolution images using SAR. The FMCW-SAR is a cheaper alternative for pulsed SAR since it requires lower transmitted power. Hence, these advantages make the FMCW-SAR a possible choice for various application\cite{navneet}. In addition, the FMCW-SAR transceiver provides a major benefit on cost-effectiveness, power consumption, volume, and weight\cite{wang2017260}.\par

The stretch processing\cite{torres},\cite{stretch},\cite{dechirp} and the matched-filter \cite{wang} are two common architectures which are widely employed in the FMCW radars. The stretch processing architecture is based on mixing the received and transmitted signal to generate a beat frequency signal. It presents high-resolution imaging with a low sampling rate which is very beneficial and cost-effective for various applications. In this architecture, to identify targets' location or ranges profile, the Fast Fourier Transform (FFT) is applied to the sampled signal. In addition to stretch processing, the matched filter architecture is also employed for FMCW radars. In this architecture, the received signal, after being down-converted, is sampled. Then, the matched filter is used to identify the targets' location or range profiles. Therefore, it would be able to achieve optimal Signal to Noise Ratio (SNR) because of using the matched filter. However, this architecture requires a high sampling rate analog to digital converter (ADC). This is a major disadvantage of this architecture and leads to an increase in the cost.\par
In this paper, a novel signal processing method is proposed. This method enables the use of a matched filter in the output of the stretch processing architecture. So, this new architecture can operate on a low sampling rate for sake of using stretch processing architecture. On the other hand, it exploits a matched filter in the output of the stretch processing, so it can achieve optimal SNR.\par

This paper is organized as follows: the FMCW radar is elaborated in section II to prepare the foundation for the next section. This section also presents the mathematical analysis for matched filter and the stretch processing architecture. In section III, the proposed radar architecture which benefits from a novel signal processing method is completely studied. Section III also gives a mathematical analysis related to the proposed architecture. Section VI evaluates the simulation results. Finally, section VI includes the conclusion.
\section{FMCW Radar}

\subsection{Stretch Processing}
\begin{figure}[b!]

\centering
\includegraphics[scale=0.42]{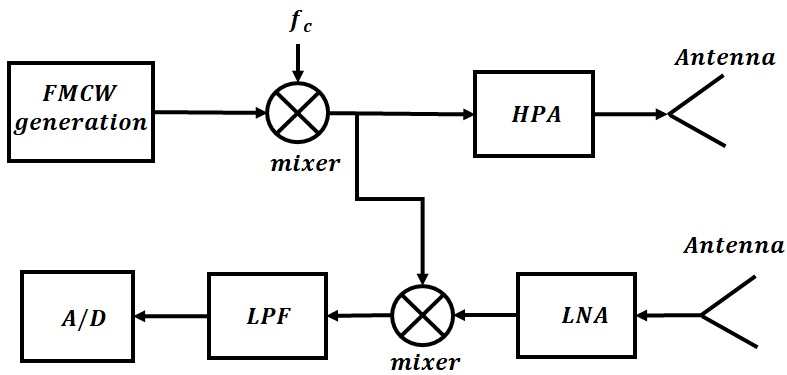}
\caption{Simplified block diagram of a stretch processing  architecture}
\label{stretch}
\end{figure}

The stretch processing architecture is widely employed in FMCW radars for sake of It's great benefits of low sampling rate and inexpensive hardware\cite{ting2017fmcw}. It also provides a fine resolution for linear frequency modulated (LFM) radars\cite{mir2015low}. The stretch processing architecture is illustrated by Fig.\ref{stretch}. As it can be seen, the stretch processing architecture is based on mixing the reflected signal off a target with a local FMCW signal. Therefore, a  beat frequency signal is generated by the mixer. The frequency of the beat frequency signal directly depends on the target range. According to Fig.\ref{stretch}, the transmitted signal is defined as:
\begin{equation}
S_t(t)=A exp\big(\phi(t)\big) .rect(\frac{t-T/2}{T})
\end{equation}
\begin{equation}
\phi(t)=\pi kt^2
\end{equation}

\(T\) is the chirp time and \(k\) is the chirp rate. The transmitted signal is reflected off a target at a range of \(R\) and the received signal can be represented as:
\begin{equation}
S_r(t)=A exp\big(\phi(t-\tau)\big) .rect(\frac{t-\tau-T/2}{T})
\end{equation}

\(\tau=2R/C\) is the time delay and \(C\) is the speed of light. Then, the received signal is mixed with a local FMCW signal in the mixer. The beat frequency signal is generated as follow:
\begin{equation}
\begin{split}
S_{if}(t) & =A S_t^{*}(t).S_r(t)=exp\big(\phi(t-\tau)-\phi(t)\big) \\
& rect(\frac{t-\tau-T/2}{T}).rect(\frac{t-T/2}{T})
\end{split}
\end{equation}
the \(S_{if}\)  consists of a beat frequency signal whose frequency  is directly proportional to the target range. Hence, the \(S_{if}\) signal can be represented as:
\begin{equation}
S_{if}(t)= A exp(\pi kt_d t-\pi kt_d)
\end{equation}

\begin{figure}[b!]

\centering
\includegraphics[scale=0.42]{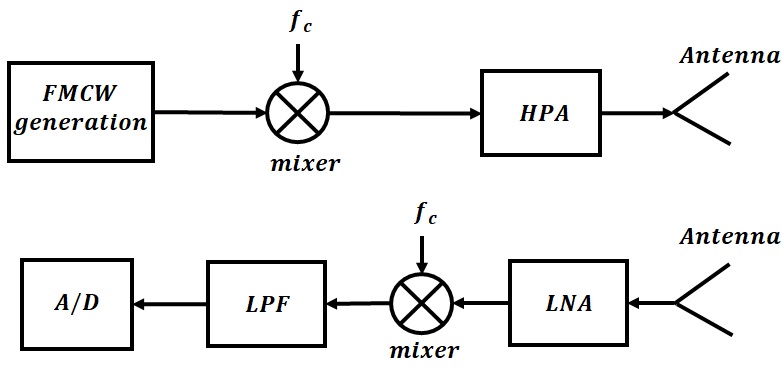}
\caption{Simplified block diagram of a matched-filter architecture}
\label{figure1}
\end{figure}

 Since the frequency of \(S_{if}\) directly depends on the target range, the Fast Fourier Transform (FFT) can precisely identify the target location. In near-range applications, the frequency of \(S_{if}\) is often low. So, it requires a low sampling rate ADC which is a great benefit of a stretch processing architecture. 
\subsection{Matched filter}
The matched filter architecture is an alternative to stretch processing architecture in FMCW radars. This architecture is widely employed in many SAR imaging algorithms such as chirp scaling, Range-Doppler, frequency scaling, etc. \cite{High-Resolution}. The matched-filter architecture is shown in Fig.\ref{figure1}. The transmitted signal is defined as:
\begin{equation}
S_t(t)=A exp\big(\phi(t)\big) .rect(\frac{t-T/2}{T})
\end{equation}
\begin{equation}
\phi(t)=\pi kt^2
\end{equation}
\(T\) is the chirp time duration  and \(k\) is the chirp rate. Likewise the stretch processing architecture, the received signal can be represented as: 
\begin{equation}
S_r(t)=A exp\big(\phi(t-\tau)\big) .rect(\frac{t-\tau-T/2}{T})
\end{equation}
In this architecture, the received signal is down-converted to baseband and then is sampled by ADC. The ADC sampling rate must be twice the bandwidth of the received signal based on the Nyquist theorem. Hence, the high-resolution application necessitates a higher sampling rate ADC which leads to an increase in the cost of radar devices. Just after the signal is sampled by ADC, the matched filter is applied to the signal to identify the target location. Thus, the matched filter architecture can achieve the optimal SNR owing to the use of the matched filter. This radar architecture is capable to operate in a noisy environment. Both stretch processing and matched filter architecture are used for SAR imaging. It should be noted that the matched filter architecture generates images with higher quality.    
\section{new matched-filter based method architecture}
In this section, a novel radar architecture based on the matched filter is proposed. This new architecture aims to operate on a low sampling rate and achieve optimal SNR. To this end, stretch processing architecture (refer to Fig.\ref{stretch}) is used to keep the sampling rate low. A matched filter is also employed in the output of the stretch processing to achieve optimal SNR. Fig.\ref{post1} illustrates the proposed architecture. In this architecture, a novel processing method is exploited (refer to Fig.\ref{post}). This method enables employing a matched filter in the output of the stretch processing. Thus, proposed radar can easily achieve optimal SNR.
\subsection{mathematical analysis}
\begin{figure}[b!]

\centering
\includegraphics[scale=0.26]{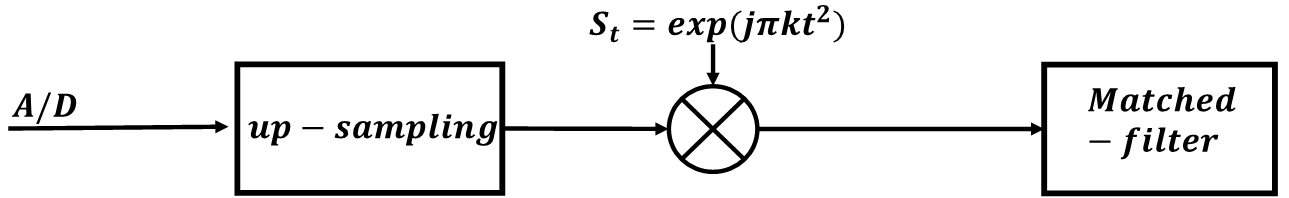}
\caption{Proposed Pre-processing section which enable stretch processing architecture to exploit matched-filter to archive optimal SNR }
\label{post}
\end{figure}

\begin{figure}[b!]

\centering
\includegraphics[scale=0.22]{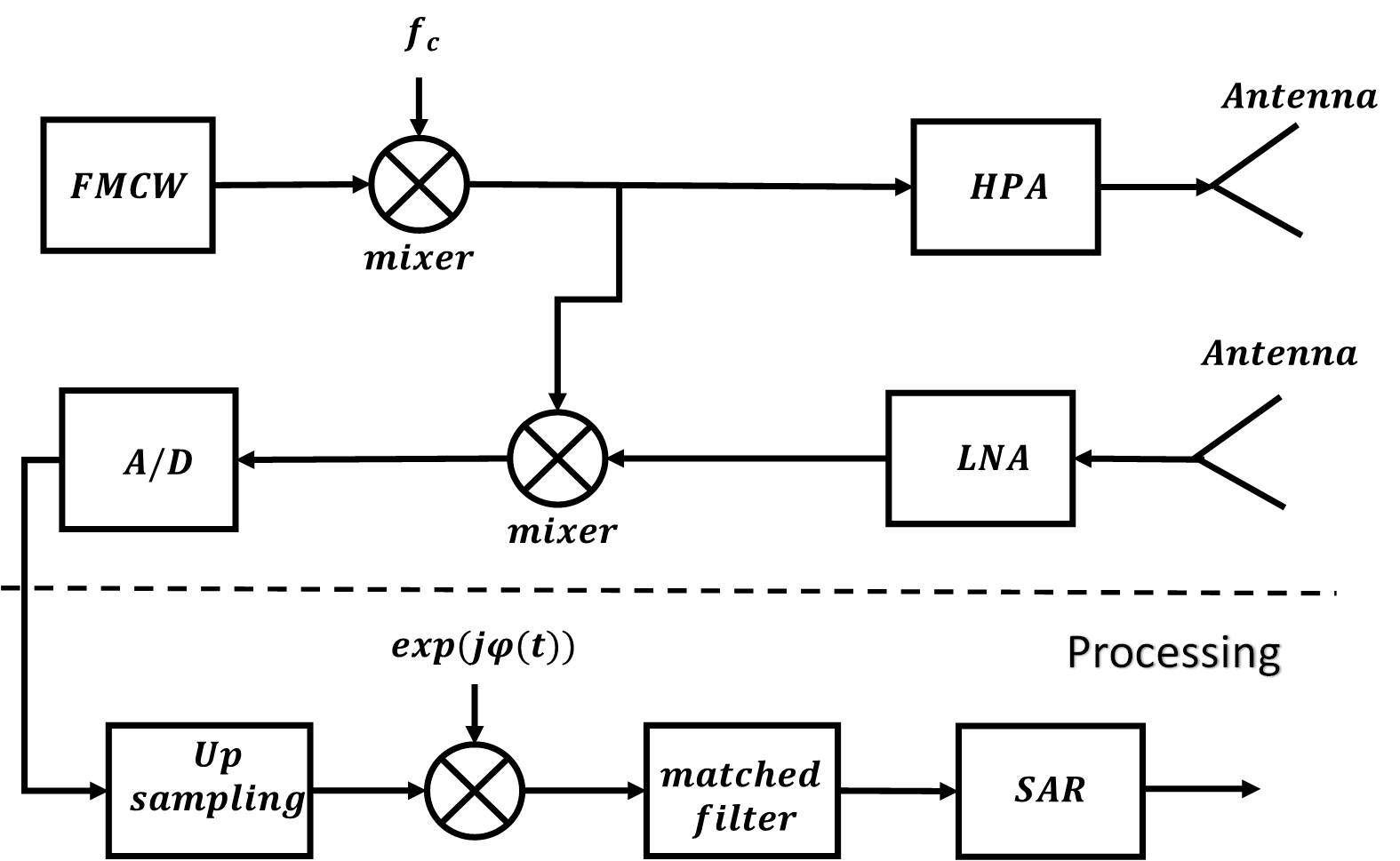}
\caption{the Proposed matched-filter based stretch processing architecture }
\label{post1}
\end{figure}

This section aims to give a mathematical analysis of the proposed radar architecture. As mentioned earlier, the proposed radar uses stretch processing in the hardware section due to lower the sampling rate. According to Fig.\ref{post1}, the signal which is sampled by ADC is expressed as:

\begin{equation}
S_{if}(t)=A S_r(t).S_t^{*}(t)
\end{equation}
According to Fig.\ref{post}, the sampled signal is up-sampled and then is multiples by \(exp(j2\pi kt^2)\). So, \(S_t^{*}(t)\) term will be eliminated from \(S_{if}(t)\) and the received signal is reconstructed as follow:

\begin{equation}
S_{rec}(t)=A S_r(t)
\end{equation}
So, the target location can be readily identified by a matched filter. Using the proposed processing method (Fig.\ref{post}) enables employing matched filter in the output of the stretch processing. Thus, proposed radar can achieve optimal SNR while benefits from a low sampling rate.

\section{simulation result}
\begin{figure}
     \centering
     \begin{subfigure}{0.4\textwidth}
         \centering
         \includegraphics[width=\textwidth]{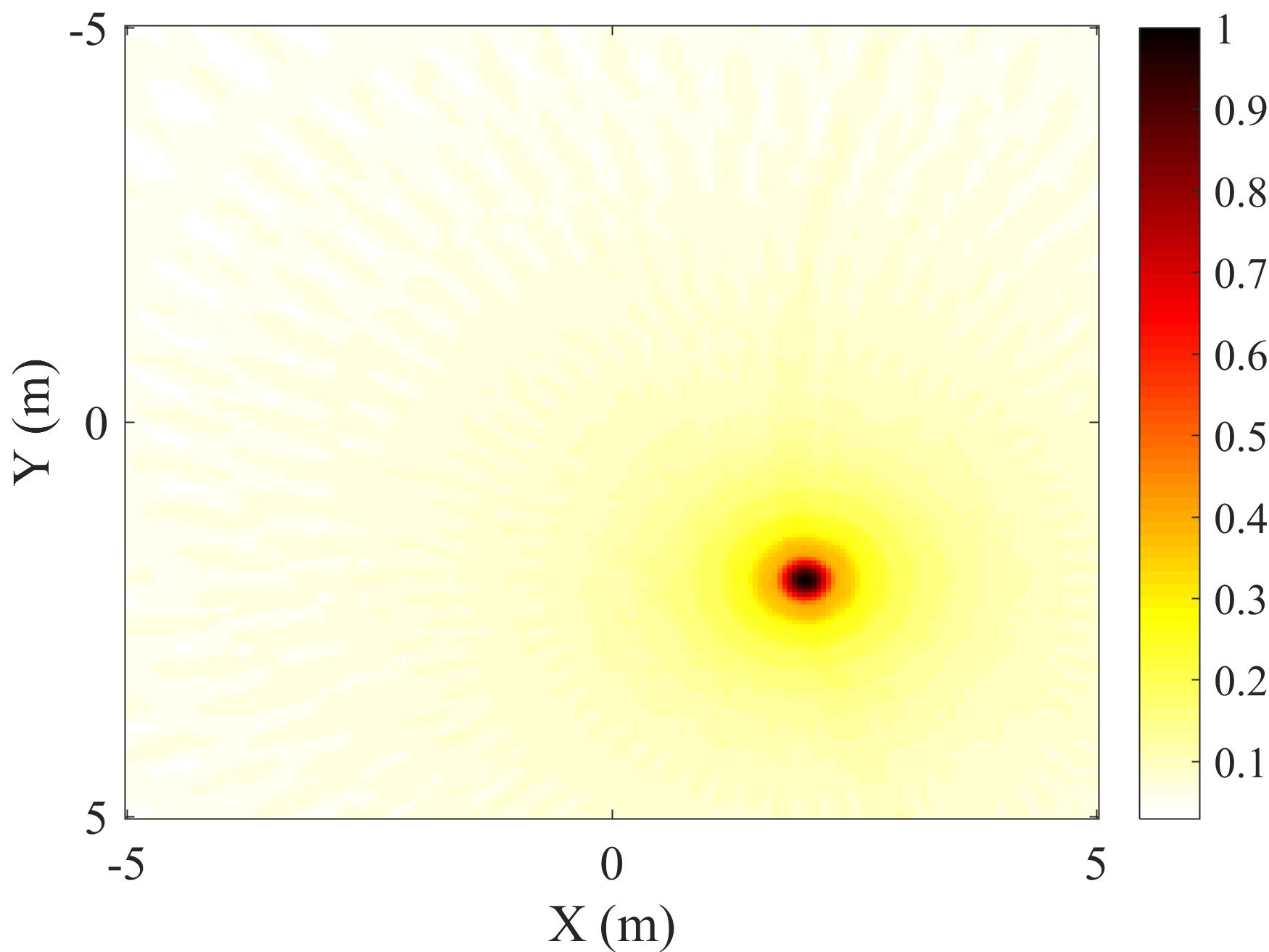}
                \caption{\(SNR_{in}=0\)}

     \end{subfigure}
     \begin{subfigure}{0.4\textwidth}
         \centering
         \includegraphics[width=\textwidth]{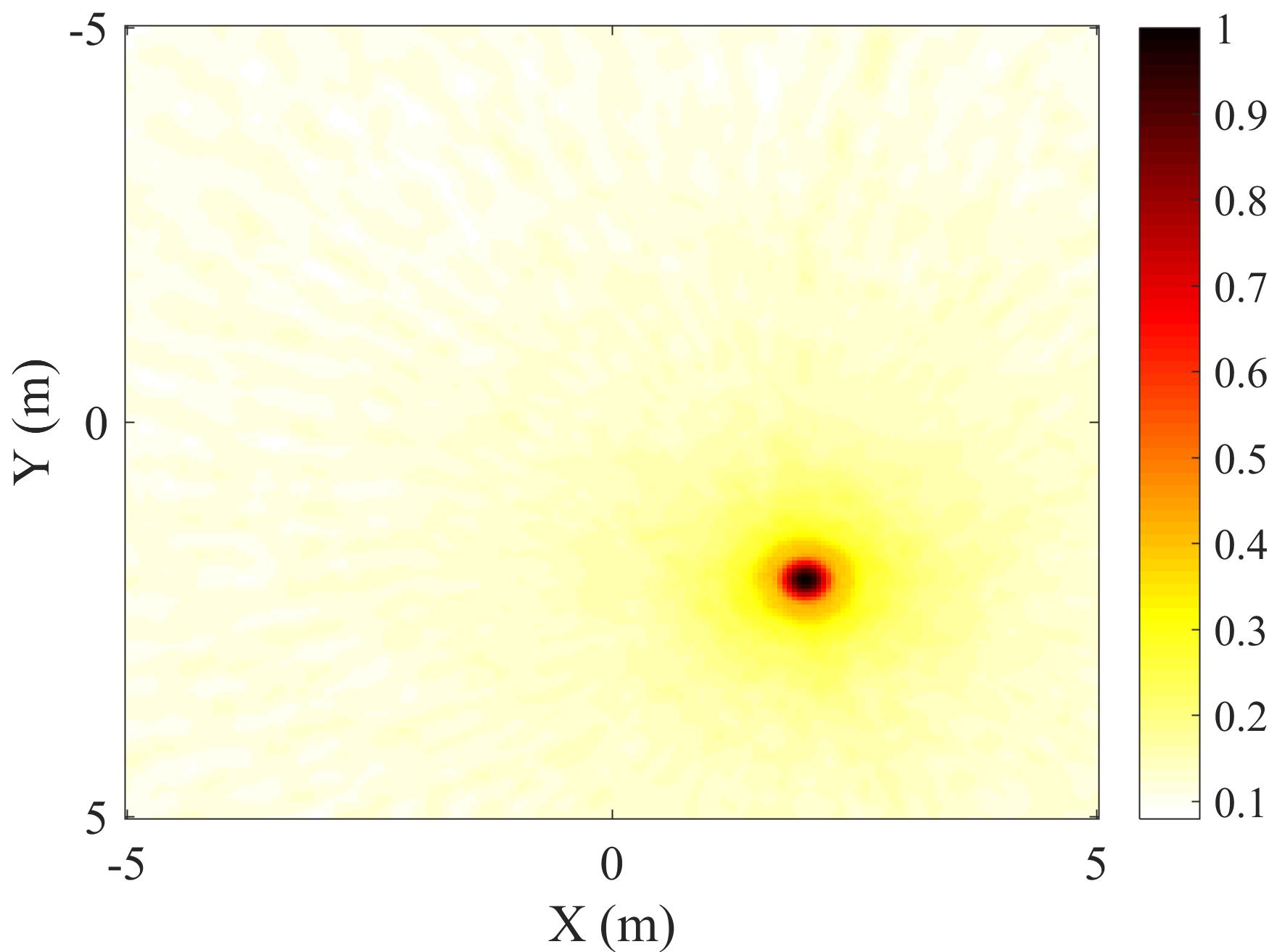}
         \caption{\(SNR_{in}=-10\)}
         
     \end{subfigure}
     \begin{subfigure}{0.4\textwidth}
         \centering
         \includegraphics[width=\textwidth]{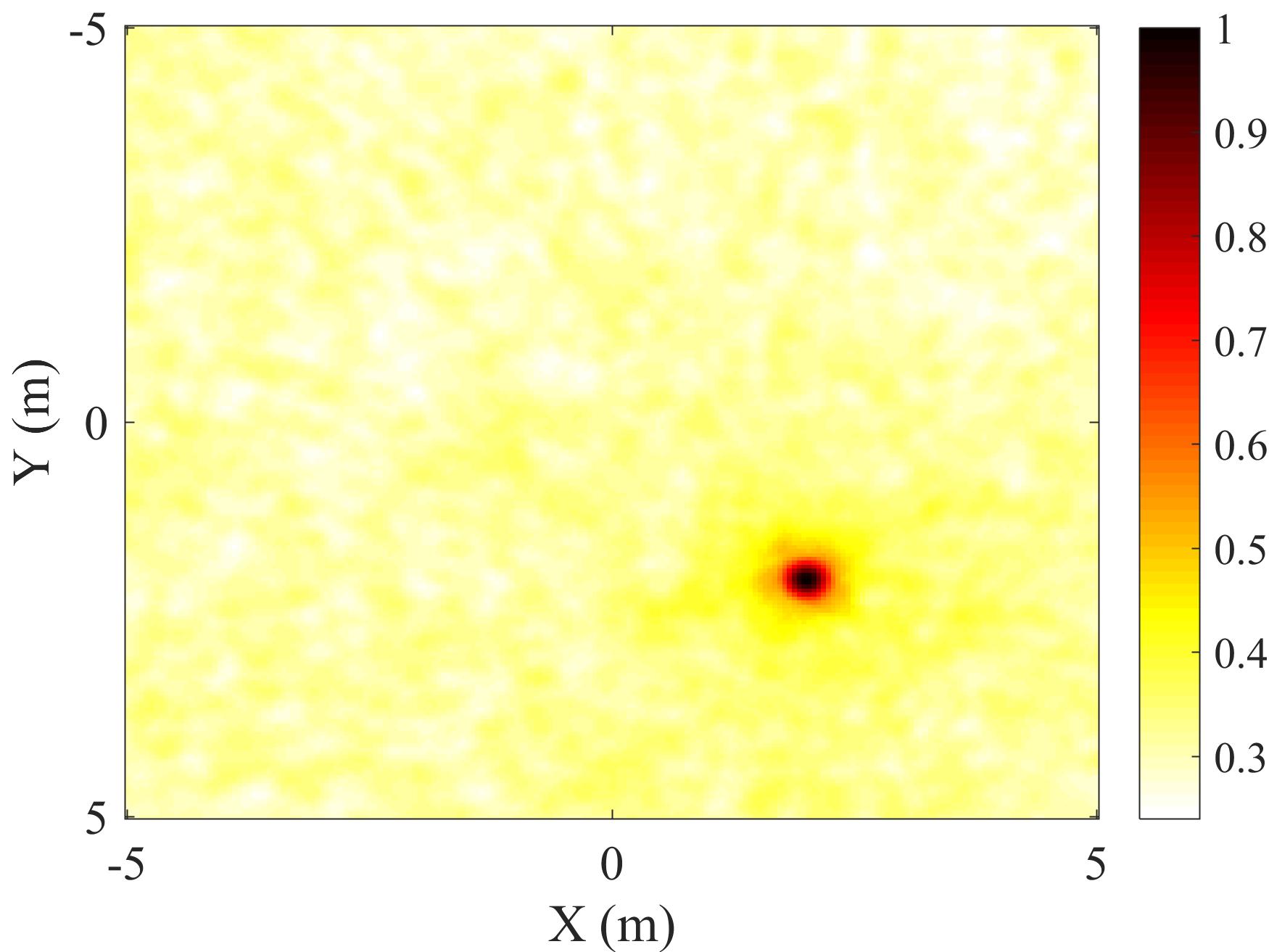}
        \caption{\(SNR_{in}=-20\)}
     \end{subfigure}

     \begin{subfigure}{0.4\textwidth}
         \centering
         \includegraphics[width=\textwidth]{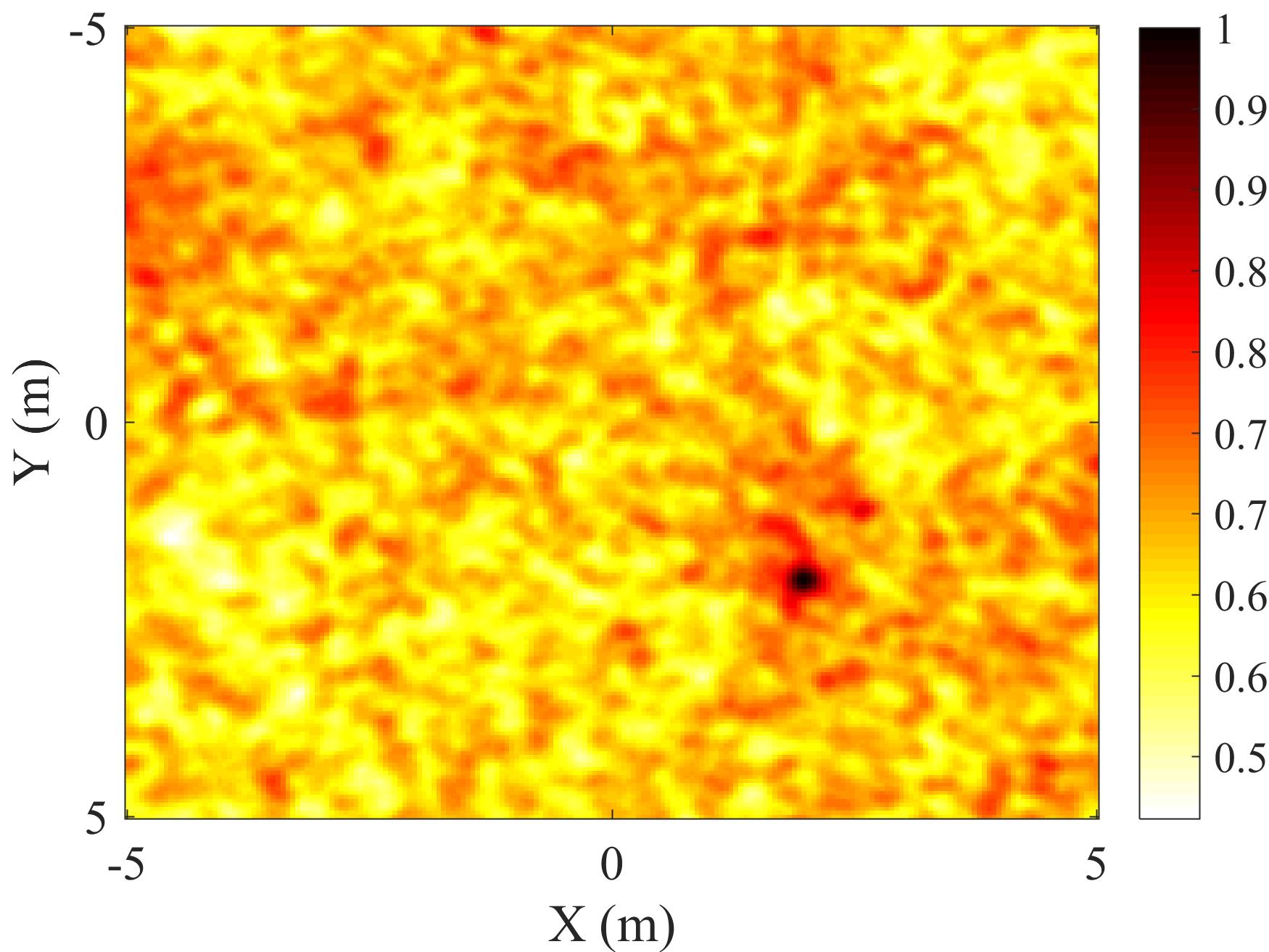}
        \caption{\(SNR_{in}=-30\)}
     \end{subfigure}
     
       \caption{The generated Images using the traditional Stretch processing architecture}
        \label{str}
\end{figure}

\begin{figure}
     \centering
     \begin{subfigure}{0.4\textwidth}
         \centering
         \includegraphics[width=\textwidth]{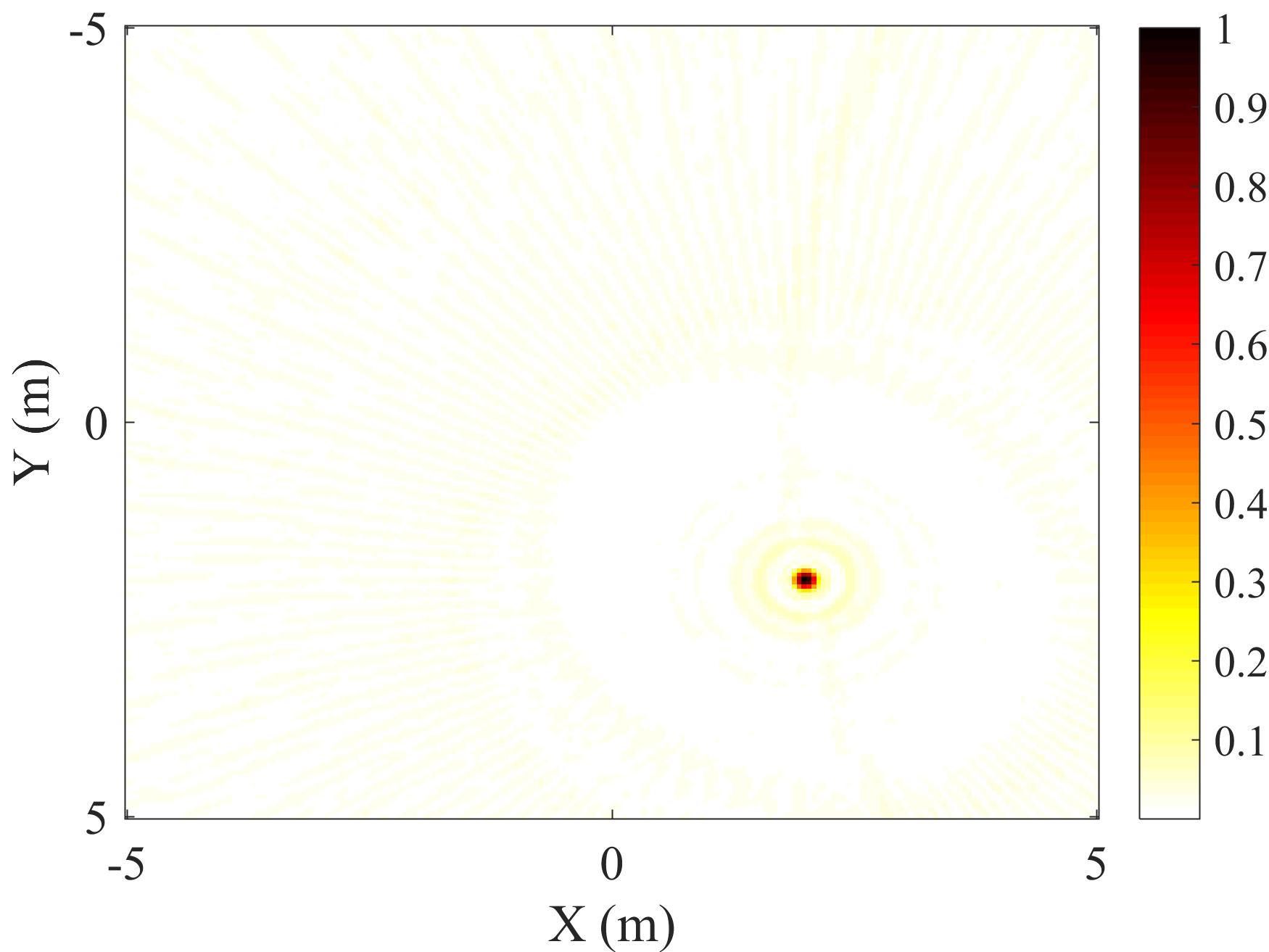}
\caption{\(SNR_{in}=0\)}
     \end{subfigure}
     \begin{subfigure}{0.4\textwidth}
         \centering
         \includegraphics[width=\textwidth]{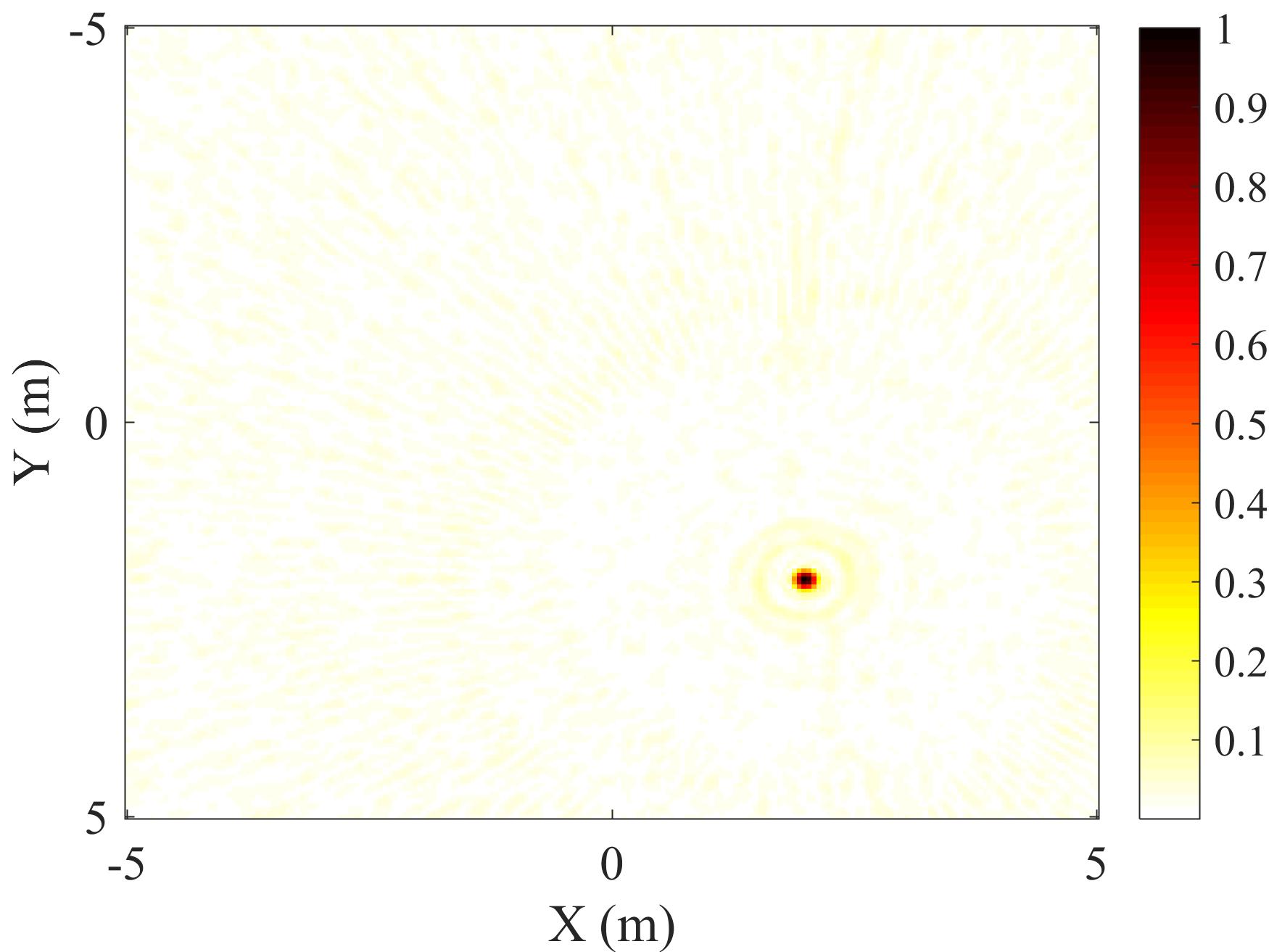}
         \caption{\(SNR_{in}=-10\)}
         
     \end{subfigure}
     \begin{subfigure}{0.4\textwidth}
         \centering
         \includegraphics[width=\textwidth]{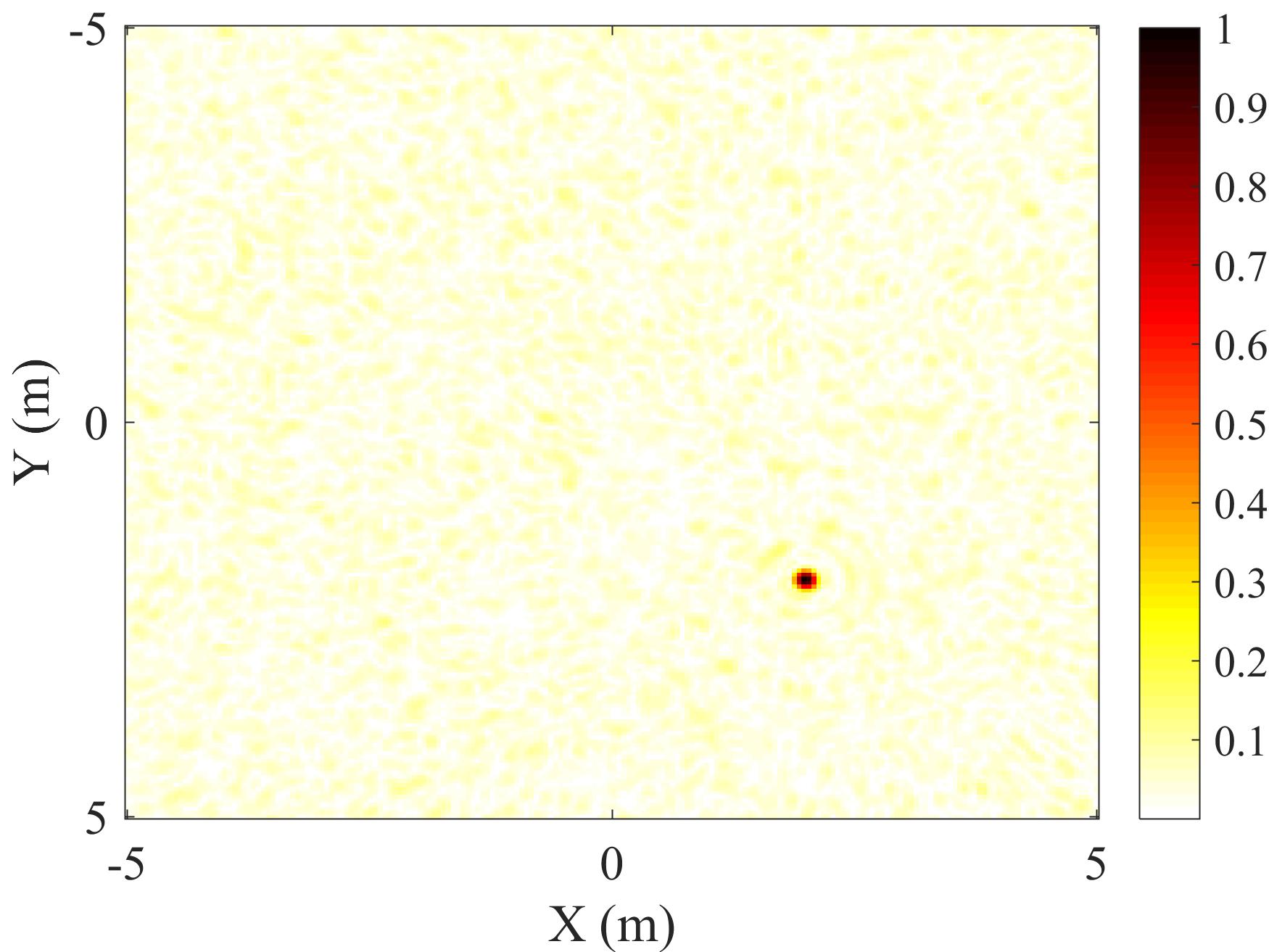}
        \caption{\(SNR_{in}=-20\)}
     \end{subfigure}

     \begin{subfigure}{0.4\textwidth}
         \centering
         \includegraphics[width=\textwidth]{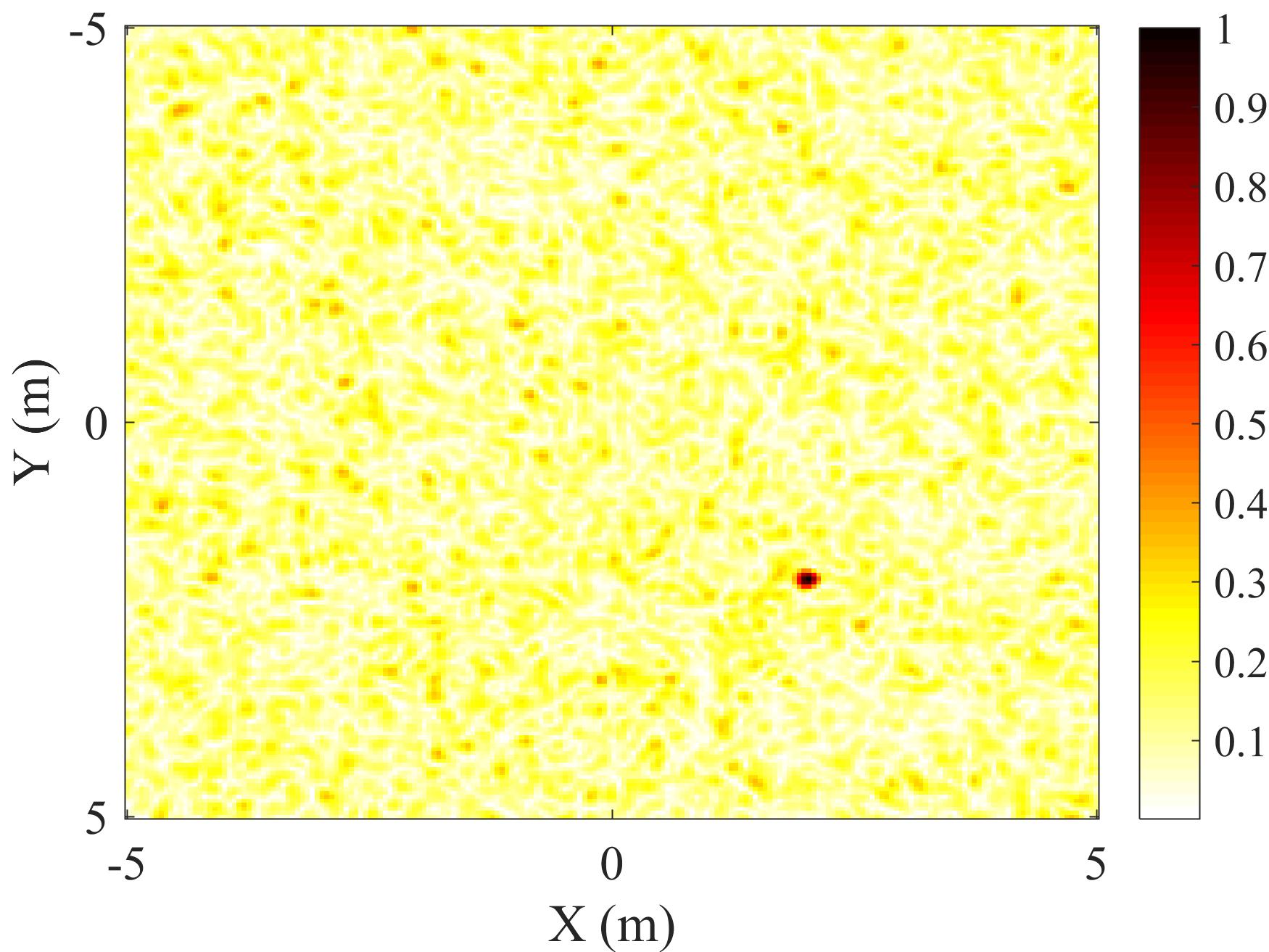}
\caption{\(SNR_{in}=-30\)}
     \end{subfigure}
     
        \caption{The generated Images using the proposed radar architecture}
        \label{mat}
\end{figure}

In this section, the performance of traditional FMCW radar and proposed radar architecture in improving SNR is fully analyzed. In this paper, the circular global back projection (GBP) algorithm\cite{oloumi} is used to generate an image. We consider four scenarios in which a target is located at \((x,y)=(2,-2)\). An FMCW radar with 1GHz bandwidth and 10\(\mu\)s chirp duration time is also utilized. Fig.\ref{str} and Fig.\ref{mat} show the images generated by two radar architectures with different SNR. As can be seen from these images, the proposed radar can generate higher quality images in comparison to the traditional stretch processing architecture.  Table.\ref{gain} also shows the compression gain of these two radars architecture in different SNR. As can be seen, the proposed radar brings a better gain over stretch processing. It is worthy to note that the compression gain of the matched filter is defined as\cite{santra}:
\begin{equation}
GP=WT
\label{GP}
\end{equation}
W and T are FMCW bandwidth and chirp time duration respectively. According to (\ref{GP}), the compression gain of a matched-filter which is applied to an FMCW signal (\(W=1 GHz\), T= 10\(\mu\)s) is about 40 dB. The compression gain value for the proposed method is higher than the analytical value (refer to Table.\ref{gain}). This additional gain (approximately 8 dB) is achieved due to the SAR algorithm. The SAR algorithm utilizes a kind of averaging method\cite{oloumi1}. In this algorithm, a radar moves along a direction and acquires range profiles. Ultimately, these range profiles are combined to generate an image. Therefore, the SAR algorithm benefits from an averaging method indirectly. \par

\begin{table}[htbp]
\caption{Compression Gain}
\begin{center}
\begin{tabular}{|c|c|c|c|}
\hline
 \(SNR_{in}\) (dB)& Matched-Filter based method (dB) &  Stretch Processing (dB) \\ 
\hline
0 & 48 & 37 \\ 
\hline
-10 & 48.2 & 34 \\ 
\hline
-20 & 48.6 & 33.8 \\ 
\hline
-30 & 49 & 33.4 \\ 
\hline
\end{tabular}

\end{center}
\label{gain}
\end{table}

\section{conclusion}
In this paper, a novel signal processing method is presented. This method enables the exploit of matched-filter in the output of the stretch processing. Thus, this new radar architecture can achieve optimal SNR while operates on a low sampling rate. The analytical analysis of the proposed radar architecture is thoroughly studied. The impact of the proposed radar architecture on improving SNR is validated. The GPB algorithm is also employed to generate images based on the SAR principle in both stretch processing and the proposed architecture. The simulation results show that the proposed architecture demonstrates a potential capability for generating an image with higher quality compared to stretch processing.

\vspace{12pt}
\bibliographystyle{IEEEtran}
\bibliography{ref.bib}

\begin{thebibliography}{10}
\providecommand{\url}[1]{#1}
\csname url@samestyle\endcsname
\providecommand{\newblock}{\relax}
\providecommand{\bibinfo}[2]{#2}
\providecommand{\BIBentrySTDinterwordspacing}{\spaceskip=0pt\relax}
\providecommand{\BIBentryALTinterwordstretchfactor}{4}
\providecommand{\BIBentryALTinterwordspacing}{\spaceskip=\fontdimen2\font plus
\BIBentryALTinterwordstretchfactor\fontdimen3\font minus
  \fontdimen4\font\relax}
\providecommand{\BIBforeignlanguage}[2]{{%
\expandafter\ifx\csname l@#1\endcsname\relax
\typeout{** WARNING: IEEEtran.bst: No hyphenation pattern has been}%
\typeout{** loaded for the language `#1'. Using the pattern for}%
\typeout{** the default language instead.}%
\else
\language=\csname l@#1\endcsname
\fi
#2}}
\providecommand{\BIBdecl}{\relax}
\BIBdecl

\bibitem{ting2017fmcw}
J.-W. Ting, D.~Oloumi, and K.~Rambabu, ``Fmcw sar system for near-distance
  imaging applications—practical considerations and calibrations,''
  \emph{IEEE Transactions on Microwave Theory and Techniques}, vol.~66, no.~1,
  pp. 450--461, 2017.

\bibitem{meta}
A.~Meta, P.~Hoogeboom, and L.~P. Ligthart, ``Signal processing for fmcw sar,''
  \emph{IEEE Transactions on Geoscience and Remote Sensing}, vol.~45, no.~11,
  pp. 3519--3532, 2007.

\bibitem{charvat2006low}
G.~L. Charvat and L.~C. Kempel, ``Low-cost, high resolution x-band laboratory
  radar system for synthetic aperture radar applications,'' pp. 529--531, 2006.

\bibitem{navneet}
S.~Navneet, A.~Roy, and C.~Bhattacharya, ``High-resolution sar image generation
  by subaperture processing of fmcw radar signal,'' \emph{IEEE Geoscience and
  Remote Sensing Letters}, vol.~11, no.~11, pp. 1866--1870, 2014.

\bibitem{wang2017260}
Y.~Wang, L.~Lou, B.~Chen, Y.~Zhang, K.~Tang, L.~Qiu, S.~Liu, and Y.~Zheng, ``A
  260-mw ku-band fmcw transceiver for synthetic aperture radar sensor with
  1.48-ghz bandwidth in 65-nm cmos technology,'' \emph{IEEE Transactions on
  Microwave Theory and Techniques}, vol.~65, no.~11, pp. 4385--4399, 2017.

\bibitem{torres}
J.~A. Torres, R.~M. Davis, J.~D.~R. Kramer, and R.~L. Fante, ``Efficient
  wideband jammer nulling when using stretch processing,'' \emph{IEEE
  Transactions on Aerospace and Electronic Systems}, vol.~36, no.~4, pp.
  1167--1178, 2000.

\bibitem{stretch}
W.~J. Caputi, ``Stretch: A time-transformation technique,'' \emph{IEEE
  Transactions on Aerospace and Electronic Systems}, no.~2, pp. 269--278, 1971.

\bibitem{dechirp}
R.~Middleton, ``Dechirp-on-receive linearly frequency modulated radar as a
  matched-filter detector,'' \emph{IEEE Transactions on Aerospace and
  Electronic Systems}, vol.~48, no.~3, pp. 2716--2718, 2012.

\bibitem{wang}
J.~Wang, D.~Cai, and Y.~Wen, ``Comparison of matched filter and dechirp
  processing used in linear frequency modulation,'' in \emph{2011 IEEE 2nd
  International Conference on Computing, Control and Industrial Engineering},
  vol.~2.\hskip 1em plus 0.5em minus 0.4em\relax IEEE, 2011, pp. 70--73.

\bibitem{mir2015low}
H.~S. Mir and U.~K.~T. Wong, ``Low-rate sampling technique for range-windowed
  radar/sonar using nonlinear frequency modulation,'' \emph{IEEE Transactions
  on Aerospace and Electronic Systems}, vol.~51, no.~3, pp. 1972--1979, 2015.

\bibitem{High-Resolution}
R.~Yang, H.~Li, S.~Li, P.~Zhang, L.~Tan, X.~Gao, and X.~Kang,
  \emph{High-Resolution Microwave Imaging}.\hskip 1em plus 0.5em minus
  0.4em\relax Springer, 2018.

\bibitem{oloumi}
D.~Oloumi, ``Oil well monitoring by ultra-wideband ground penetrating synthetic
  aperture radar,'' 2012.

\bibitem{santra}
A.~Santra, R.~Srinivasan, K.~Jadia, and G.~Alleon, ``Ambiguity functions,
  processing gains, and cramer-rao bounds for matched illumination radar
  signals,'' \emph{IEEE Transactions on Aerospace and Electronic Systems},
  vol.~51, no.~3, pp. 2225--2235, 2015.

\bibitem{oloumi1}
D.~Oloumi, J.-W. Ting, and K.~Rambabu, ``Design of pulse characteristics for
  near-field uwb-sar imaging,'' \emph{IEEE Transactions on Microwave Theory and
  Techniques}, vol.~64, no.~8, pp. 2684--2693, 2016.

\end{thebibliography}
\end{document}